*Many Worlds? Everett, Quantum Theory and Reality*
Edited by; Simon Saunders, Jonathan Barrett, Adrian Kent and David Wallace
Oxford University Press,  ISBN: 9780199560561; £55 hardback; pp. xvi + 618.

*1. An anniversary*
2007 saw the fiftieth anniversary of Everett's proposal (1957) for how to solve quantum theory's notorious 'measurement problem'. To commemorate it, two conferences were held: at Oxford University, UK, and the Perimeter Institute for Theoretical Physics, Canada. In this book, the organizers have collected most of the papers, by physicists as well as philosophers, which were presented; together with some other papers and discussions.

I should begin by describing the measurement problem and standard quantum theory's response to it. The measurement problem ('Schroedinger's cat paradox') is the threat that the lack of values for physical quantities such as position, momentum and energy, which is characteristic of quantum theory's description of microscopic systems such as electrons, should also infect the familiar macroscopic realm of tables, chairs---what J.L.Austin called 'medium-sized dry goods'---with their apparently definite values for position, momentum etc. The threat is clearest if one considers a measurement situation. Quantum theory apparently predicts that measuring, for example, the momentum of an electron, when it is in a state that is not definite for momentum, (a 'superposition of momentum eigenstates') should lead to the pointer of the apparatus having no definite position---it should be in a superposition of position eigenstates.

By about 1935, standard quantum theory had settled on the following response to this problem. One postulates that the quantum state of both the microscopic system and the apparatus changes discontinuously after the measurement interaction, so that the apparatus' pointer gets a definite position. (This postulate is called the 'projection postulate'; the change of state is called, more colloquially, the 'collapse of the wave-packet'.) This is of course unsatisfactory, since 'measurement' is vague. And there is worse trouble: the projection postulate contradicts quantum theory's usual law of how states change over time, viz. the Schroedinger equation, which prescribes a continuous and deterministic evolution.

Everett proposes that one can solve the measurement problem without any recourse to the projection postulate. In brief: he claims that the universe as a whole has a quantum state, which always evolves according to the Schroedinger equation. Agreed: the measurement problem suggests this state will be a superposition corresponding to many different definite macroscopic realms ('macrorealms'). But Everett suggests that we should explain our experience of a single definite macrorealm, by postulating that all the various definite macrorealms are actual. Thus the universe (what philosophers nowadays call 'the actual world') contains a plethora of Everettian 'worlds', where each such 'world' is something like the familiar macroscopic realm, with all tables, chairs etc. in definite positions. But the worlds differ among themselves about these positions, i.e.

about where the various macroscopic objects are; and we just happen to be in one world rather than any of the others. Hence this book's title.

This dizzying vision obviously calls out for philosophical clarification, since it involves central metaphysical topics such as possibility and persistence through time. Indeed, it also bears on the relation between mind and matter. For some versions of Everett's proposal envisage an ontology of many `minds'; i.e. they claim that to each sentient brain (a human's, a cat's …) there corresponds a plethora of minds (or if you prefer, mental states): their experiences differing about such matters as the location of macroscopic objects.

Since Everett's original proposal in 1957, a philosophical literature about it has grown up: especially since the 1980s, with the growth of philosophy of physics (Butterfield (1995, 2000) were two of my own attempts). But it still remains controversial, and indeed, not precisely defined: 'the Everett interpretation' remains a vague definite description. All the more reason, then, to scrutinize the various formulations, and to thoroughly assess them.

In this excellent book, some of today's most gifted practitioners—some in physics, and some in philosophy---undertake this task. Apart from the conferences' papers, including three replies and three transcripts of discussions (about ten pages each), the book also contains a few other invited papers; and a long, very helpful introduction by Saunders.

The standard is high. All the papers are worth reading, most are worth studying, and some will last as major contributions to the literature. Besides, approximately the same number of papers (or of pages) is devoted to advocacy of the Everett interpretation, as to criticism of it. So the book is even-handed, and a 'must-read' for anyone keen on, or sceptical of, the interpretation. Indeed, of the four editors, two are Everettians (Saunders and Wallace), one (Kent) is opposed, and the fourth (Barrett) is agnostic. The book also contains some discussions of rival interpretations of quantum theory, especially the pilot-wave approach of de Broglie and Bohm. For the most part, these discussions compare the ontologies or world-pictures of Everett and these rivals. This wide coverage has one significant omission: the 'many minds' version of the Everett interpretation is hardly discussed. So I recommend anyone interested to study what is surely its most sustained and detailed defence, namely by Matthew Donald (e.g. his 1997).

I cannot here report in detail even a few of the book's papers. I shall instead expound why in the last twenty years the Everett interpretation has become a leading approach to understanding quantum theory. I will emphasize philosophical issues, not physical ones. So I will set aside how efforts, from the 1970s onwards, to develop a quantum cosmology lent support to the Everett interpretation.

As I see matters, there have been three main developments: three reasons why the Everett interpretation has moved centre-stage in quantum philosophy. As we will see, they correspond to the first four of the six Parts of this book. Thus Parts 1 and 2 concern ontology, with Part 1 advocating the Everettian ontology, and Part 2 criticizing it. Similarly, Parts 3 and 4 concern probability, with Part 3 advocating the Everettian account of probability and Part 4 criticizing it.

In what follows, I must skip over Parts 5 and 6: regretfully, since they are interesting. Suffice it to say that the papers in Part 5 advocate or criticize some rivals to the Everett interpretation, such as the pilot-wave approach. Part 6 broadens the discussion.

For example, it includes a striking appeal by Deutsch, a very innovative advocate of the Everett interpretation, to explore the new physics that it promises to contain. It also includes a fascinating history of how Everett came to write his paper: viz. as a summary of his Princeton PhD under John Wheeler, who was unduly anxious to tone it down, so as to placate his---and most other physicists'---guru, Niels Bohr.

*2. A bluff?*
Before embarking on these three main developments, it will be helpful to describe why the Everett interpretation was widely regarded until about 1990 as obscure and-or inadequate, even to the point of being condemned as a mere bluff.

Recall Everett's two main claims. Quantum theory can be interpreted:

(i) with no funny business about a measurement process inducing the quantum state to "collapse" indeterministically, according to which of the alternative outcomes occurs; *and*

(ii) with no theoretical posits supplementing the state (traditionally called "hidden variables") so as to represent which outcome occurs---or indeed to represent any other physical fact.

So Everett's view is that the deterministic Schroedinger equation is always right, in the sense that the quantum state of an isolated system always evolves in accordance with it. And the quantum state 'is everything' in the sense that values are assigned to physical quantities only by the orthodox rules. In particular, no quantity is preferred by being assigned, in every state, a value; as is proposed, for the quantity position, by the pilot-wave approach. But to reconcile this uncollapsed and un-supplemented quantum state with the apparent fact that any quantum experiment has a single outcome, Everett then identifies the Appearances—our apparent macroscopic realm, with its various experiments' outcomes---with one of a vast multiplicity of realms. These are often called 'branches' rather than 'worlds'.

But however sympathetic one might feel to this dizzying vision, one naturally asks for a precise and general definition of a 'branch'. From the 1960s to the 1980s, Everettians usually defined 'branch' in terms of the pointer-quantity of a measurement-apparatus. So, rather like the pilot-wave approach, there was a preferred quantity with a definite value, albeit relative to a branch. But this sort of definition was not general enough, since there would no doubt be an apparently definite macroscopic realm, even if no experiments were ever performed, or no measurement-apparatus ever existed.

Agreed, this lacuna was understandable, since formulating a truly satisfactory definition would require one to consider all the various aspects of the "emergence" of the classical physical description of the universe. But it seemed that as long as the lacuna remained unfilled, the Everett interpretation was at best a vague promissory note.

Kent (1990) is a fine example of this sort of critique. Another influential voice was that of John Bell, whose work on quantum non-locality, and the experiments they inspired, did so much to make the rival interpretations of quantum theory, and more generally the foundations of physics, a respected topic within physics.

For example, in Bell's masterly introduction to interpreting quantum theory, he endorses the accusations of obscurity and vagueness, saying that the Everett interpretation 'is surely the most bizarre of all [quantum theory's possible interpretations]' and seems 'an extravagant, and above all extravagantly vague, hypothesis. I could almost dismiss it as silly' (1986, pp. 192, 194). Agreed, the Everett interpretation is not the only target of Bell's accusations of obscurity and vagueness. He has similar doubts about two others in his list of six possible interpretations (which he calls 'possible worlds'): namely, Bohr's complementarity interpretation, and the idea that consciousness induces the collapse of the quantum state (1986, pp. 190, 191, 194).

Hence Bell favoured two other 'worlds': supplementing the quantum state, as in the pilot-wave approach, or revising the Schroedinger equation, so as to describe the collapse of the quantum state in detail, as a physical process. As he puts it elsewhere: 'either the wave-function, as given by the Schroedinger equation, is not everything or it is not right' (1987, p. 201).

*3. Three developments*
But since 1990, Everettians have made three major improvements to their interpretation. At first glance, the first is a matter of physics, the second a matter of philosophy, and the third a matter of decision theory applied to physics. But on inspection, and as is evident when reading this book: these contrasts between physics and philosophy are superficial---a happy upshot for philosophers of physics. (Sad to say: John Bell died in 1990, so that his writings do not engage with these three developments.)

The first development is the theory of decoherence. Although the fundamental ideas were established in the early years of quantum theory (and were clear to *maestros* such as Heisenberg, Bohm and Mott), detailed models of decoherence were only developed from about 1980. In this book, this is represented by three distinguished contributors to this field of physics: Hartle, Halliwell and Zurek. (The last also writes about his approach to quantum probability; cf. Section 6 below).

The second development is the application to the problems of quantum ontology, especially Schroedinger's cat, of the philosophical idea that the objects in 'higher-level' ontology, e.g. a cat, are not some kind of aggregate (e.g. a mereological fusion) of lower-level objects, but rather dynamically stable patterns of them. This is here represented mainly by the opening paper by Wallace, who credits e.g. Dennett (1991) for the general philosophical idea. Both the idea and its application to quantum ontology are discussed in several other papers: especially the critical replies by Maudlin and Hawthorne.

The third is the development of various arguments justifying the form of quantum probabilities. I will emphasise one such line of argument. It was initiated by Deutsch in 1999, and developed by Wallace from 2002 onwards, and is nowadays often called 'the Deutsch-Wallace programme'. In this book, this line and-or some related proposals are discussed by Wallace, Saunders, Greaves and Myrvold, and Papineau; with critical replies by Albert, Price and Kent. I should add that by comparing Kent's paper here with his earlier (1990) critique of the Everett interpretation, the reader can get a good sense of all three developments. For Kent's paper here has the merits of considering not just probability, but also the first two developments; and of considering in detail several other papers in the volume.

I shall say a bit about each of these developments, in turn. But owing to lack of space, and in order to give the Everettians due credit, I shall only report the Everettian proposal. So I must ignore the various criticisms that can be made---which, to repeat, are also well represented in this book. I think this restriction is fair, in that even the critics would agree that with these three developments, the Everettians have made great strides towards rebutting the traditional accusations of obscurity or inadequacy---and so towards having an eminently tenable interpretation.

*4. Decoherence*

'Decoherence' means, in this context, the 'diffusion of coherence'. This is the fast and ubiquitous process whereby, for appropriate physical quantities, the interference terms in probability distributions, which are characteristic of the difference between a quantum state (a 'superposition') and a classical state (a 'mixture'), diffuse from the system to its environment.

In a bit more detail: at the end of the process, the quantum probabilities for any quantity defined on the system are as if the system is in one or other of a definite set of states. In many models of how a system (such as a dust-particle) interacts with its environment (such as air molecules), this set consists of coherent states. These are states which, considered as probability distributions, are sharply peaked for both position and momentum; so that a system in any such state is presumably nearly definite in both position and momentum. (But the distributions have enough spread so as to obey Heisenberg's Uncertainty Principle, which vetoes simultaneous precise values for position and momentum.)

For our purposes, decoherence has two important features: one positive, one negative. These features, and their consequences for quantum ontology and probability (discussed below) were emphasised by Saunders in the 1990s.

The positive feature is flexibility. Thus we expect the classical physical description of the world to be vindicated by quantum theory—but only approximately. Only some subset of quantities should have definite values. And maybe that subset should only be specified contextually, even vaguely. And maybe the values should only be definite within some margins of error, even vague ones. Decoherence secures this sort of flexibility. For the selection of the quantity that is preferred in the sense of having definite values (relative to a branch) is made by a dynamical process---whose definition can be legitimately varied in several ways. Three examples: the definitions of the system-environment boundary, and of the time at which the interaction ends, and of what counts as a state being 'sharply peaked' for a quantity, can all be varied.

The negative feature is that decoherence does not just by itself solve the measurement problem. More precisely: it does *not* imply that the system is in one of the set of states (typically coherent states). It implies only that the quantum probabilities are *as if* the system were in one. Furthermore, the theory implies that the system is in fact *not* in one of those states (on pain of contradicting the original hypothesis that the total system-plus-environment is in a superposition, not a mixture). To put it vividly, in terms of Schroedinger's cat: at the end of the decoherence process, the quantum state still describes two cats, one alive and one dead. It is just that the two cats are correlated with very different microscopic states of the surrounding air molecules. For example: an air molecule will bounce off a wagging upright tail, and a stationary downward one, in different directions!

*5. Patterns*

This brings us to the second development. It snatches victory from the jaws of defeat: the defeat at the end of Section 4, that decoherence apparently does not by itself solve the measurement problem. The idea is to appeal to the philosophical view that an object like a cat is *not* some kind of aggregate of microscopic objects, but rather a dynamically stable pattern of a special type---which type being spelt out by what we believe about cats (by our 'theory of cats'). Of course, this view is often associated with "functionalism" in the philosophy of mind.

Today's Everettians---or at least some prominent ones, such as Saunders and Wallace---maintain that with this second idea, we escape the quandary at the end of Section 4. That is: the final quantum state's describing two cats, one alive and one dead, is a matter of the state encoding two patterns---and the description is entirely right.

This becomes a bit clearer if we adopt a specific representation of the quantum state, for example position. Then, roughly speaking: the final state is a wave-function on the cat's classical configuration space, with two peaks: one peak over some classical configurations corresponding to a perky cat, e.g. with a wagging upright tail, the other peak over some classical configurations corresponding to a dead cat, e.g. with a stationary downward tail. But in that case: according to the idea of cats as patterns, the quantum state does indeed represent *two* cats.

In other words: we see that we *should* take the measurement problem to be solved by exactly what decoherence secures: a final state describing a living cat and a dead one. In brief: the philosophical idea of higher-level objects as patterns vindicates the Everettian vision of a multiplicity of objects.

It is worth stressing (as Wallace, for one, does) that this line of thought is independent of quantum theory's details; and so it is also independent of its various weird features (e.g. non-locality). The point is closely analogous to one which we all unhesitatingly endorse for several other physical theories. Namely, theories in which states can be added together to give a sum-state, while the component states are dynamically isolated, or nearly so (i.e. do not influence each other). Examples include the theory of water-waves, or electromagnetism. So, says Wallace, we should also endorse it in quantum theory, and accept that there are two cats.

For example: the water in Portsmouth harbour can get into a state which we describe as, e.g. a wave passing through the harbour's centre heading due West; or into a state which we describe as a wave passing through the centre heading due North; or into a state which is the sum of these. But do we face a 'Portsmouth water paradox'? Do we agonize about how the Portsmouth harbour water-system can in one place (viz. the harbour's centre) be simultaneously both Westward and Northward? Of course not. Rather, we say that waves are higher-level objects, patterns in the water-system; and that there are two waves, with the contrary properties, one Northward and one Westward. Similarly for the electromagnetic field in a certain region, and e.g. pulses of laser light travelling in different directions across it.

And similarly, says Wallace, about the quantum wave-functions defined on the classical configuration space. There is a state with two cats, one alive and one dead. And of course, there are also myriad other states, the vast majority of which do not represent macroscopic objects (patterns!) which we might recognize (as cats or as dogs or as puddles or as mud or …)---or even sums of these.

*6. Probability*
Finally, I turn to probability. As I announced, I will here confine myself to the Deutsch-Wallace programme for justifying the form of quantum probabilities (called 'Born-rule' probabilities, after Max Born who in 1927 first stated the probabilistic interpretation of the quantum state, viz. for position probabilities). This may sound, to someone not immersed in quantum philosophy, a very arcane topic. But in fact, it engages closely with familiar central issues in philosophy of probability, like chance, credence and the relations between them, such as David Lewis' Principal Principle (1980).

To explain this, I will first clarify how the Everett interpretation faces two apparent problems about probability: a qualitative one and a quantitative one. I will discuss these in turn. I should also warn the reader that, since I must be brief, my discussion will elide some distinctions, for example about kinds of uncertainty, which some of this book's authors hold dear!

The qualitative problem is that probability seems to make no sense, if all possible outcomes of a putatively probabilistic process in fact occur: as the Everettian says they do, at least for quantum measurements and the other processes such as radioactive decay, traditionally considered as indeterministic collapses of the quantum state. For recall that according to the Everettian, the quantum state always evolves deterministically, so that during such a process, it evolves to include a term, i.e. a summand in the sum, for each outcome.

Nowadays, the main Everettian answer to this problem---both in this book, and elsewhere---is to invoke subjective uncertainty. The idea is an analogy with how probability is taken as subjective uncertainty, for a deterministic process of the familiar classical kind. For such a process, a unique future sequence of states is determined by the present state (together with the process' deterministic law). But the agent or observer does not know this sequence in advance, either because she does not know the present state in full detail or because it is too hard to calculate from it the future sequence. Similarly, says the Everettian: probability can be taken as subjective uncertainty, for a deterministic process of the unfamiliar Everettian kind. For such a process, a unique future sequence of 'global' states is again determined by the present quantum state (together with the Schroedinger equation). But here, one can assume the agent or observer *does* know the present state, *and* how to calculate from it the future sequence. That is: the agent or observer is nevertheless uncertain since, thanks to the impending 'branching' or 'splitting', she will not experience any such future 'global' state, i.e. she will not experience the outcomes corresponding to all its terms. At each future time, she will only experience one outcome---and is thus uncertain about which. Thus this kind of uncertainty, compatible with full knowledge of the global state and the laws, is rather like the self-locating uncertainty discussed by philosophers under the heading 'the essential indexical' (e.g. Perry 1979, Lewis 1979).

So much by way of a brief statement of the Everettians' answer to the qualitative problem. (As I warned at the start of this Section: my phrase 'rather like' papers over a debate between some of this book's authors about the nature of this uncertainty.) I turn to the quantitative problem.

We can introduce this by again imagining that a quantum system is subjected to a sequence of measurements. Then according to the Everettian, the quantum state evolves over the course of time so as to encode all possible sequences of outcomes: formally, it has a term (i.e. a summand in the sum) representing each sequence of outcomes. For example, consider a toy-model in which there are ten measurements, each with two outcomes (H and T say!). Then there are $2^{10} = 1024$ sequences of outcomes; and so the Everettian will say there are 1024 terms in the quantum state.

Since according to the Everettian, each such sequence actually occurs, it seems at first that the probability of a sequence should be given by the naïve 'counting measure': each sequence has probability 1/1024. And so more generally, it seems that the probability of an event corresponding to a set of sequences, such as three of the ten measurements having outcome H, is the sum of the elementary probabilities of its component sequences. But this amounts to assuming that the two outcomes H and T are equiprobable (and that the measurements form independent trials in the sense of probability theory). And this spells disaster for the Everettian: the counting measure probabilities bear no relation to the quantum Born-rule probabilities, and so 'counting worlds' seems to conflict with quantum theory's treatment of probability.

Today's Everettians have a twofold answer to this. The first part is to point out that decoherence, thanks to its flexibility, refutes the toy-model with its naïve counting measure. (This is emphasized by Saunders; cf. Section 4.) That is: on any precise definition of 'branch' for the systems concerned, there will be trillions of branches, wholly independently of the number of kinds of outcome registered by the measurement apparatus (in my example: just two, H and T). And more important: because decoherence is vague, there is *no* definite number of branches which we need to—or could!---appeal to in order to give an account of probability. In short: the naïve counting measure is a chimera and a canard---to be rejected out of hand.

The second part is what I have labelled as the Deutsch-Wallace programme. This builds on the previous Everettian answer to the qualitative problem, i.e. the invocation of subjective uncertainty. Recall the tradition, in subjective decision theory, of representation theorems to the following effect: an agent whose preferences for gambles (encoding certain degrees of belief and certain desires) conform to a certain set of axioms, which look to be rationally compelling, must have degrees of belief that are represented by a probability function. (Such theorems go back to authors such as Ramsey, de Finetti, Savage and Jeffrey.)

There is a lot to say, both technically and philosophically, about such theorems. But for our purposes, we need only note that these theorems do not dictate a specific probability function. This is of course as one would expect: surely, rationality should not dictate specific degrees of belief in arbitrary propositions!

But Deutsch and Wallace prove theorems with precisely this feature, about the specific scenario of making gambles on the outcomes of quantum measurements. And the probability function that is dictated by their axioms (which, as in the tradition, look to be rationally compelling) is precisely the orthodox Born-rule of quantum theory!

A bit more precisely: Deutsch and Wallace show that an Everettian agent who is about to observe a sequence of quantum measurements, who also knows the initial state of the quantum system to be measured, and who is forced to gamble on which outcomes she will see (in the Everettian sense of 'splitting'), and whose gambles are subject to certain rationality axioms---must apportion her degrees of belief (as shown by her betting behaviour) in accordance with the Born-rule.

To sum up: we have here an argument to the effect that, *pace* the above objection to the naïve 'counting measure', the Everettian framework not only accommodates, but even *implies*, the Born rule.

Even from this brief and vague statement, it is clear that these representation theorems are very remarkable: one might say, amazing! Indeed, they are remarkable, both technically and philosophically, and are a gold-mine for the philosophy of probability: a gold mine whose first seams are worked out in Deutsch's and Wallace's papers and the ensuing literature---including the papers, *pro* and *con*, in this book.

*7. Summary*
To sum up: I hope to have conveyed how the Everett interpretation is full of interest for philosophy. And this is not just because the original vision of 'branching' or 'splitting' obviously calls out for clarification in relation to topics such as ontology and probability. Also, and more important: the three developments of the last twenty years have both substantially improved the Everett interpretation and connected it in richer detail with such topics. Besides, the state of play about all these developments is conveyed very well by this book's high-quality discussions.

Thus I recommend the book wholeheartedly not just to any philosopher of physics, but to any metaphysician and epistemologist who is minded to have their views moulded by the deliverances of empirical enquiry. All future work on the Everett interpretation begins here.

*Acknowledgements*: I am very grateful to Matthew Donald, Anthony O'Hear, Adrian Kent, Simon Saunders and David Wallace for comments on earlier drafts: if only I could have consistently followed all suggestions!

9of the quantum system to be measured, and who is forced to gamble on which outcomes she will see (in the Everettian sense of 'splitting'), and whose gambles are subject to certain rationality axioms---must apportion her degrees of belief (as shown by her betting behaviour) in accordance with the Born-rule.

To sum up: we have here an argument to the effect that, *pace* the above objection to the naïve 'counting measure', the Everettian framework not only accommodates, but even *implies*, the Born rule.

Even from this brief and vague statement, it is clear that these representation theorems are very remarkable: one might say, amazing! Indeed, they are remarkable, both technically and philosophically, and are a gold-mine for the philosophy of probability: a gold mine whose first seams are worked out in Deutsch's and Wallace's papers and the ensuing literature---including the papers, *pro* and *con*, in this book.

*7. Summary*
To sum up: I hope to have conveyed how the Everett interpretation is full of interest for philosophy. And this is not just because the original vision of 'branching' or 'splitting' obviously calls out for clarification in relation to topics such as ontology and probability. Also, and more important: the three developments of the last twenty years have both substantially improved the Everett interpretation and connected it in richer detail with such topics. Besides, the state of play about all these developments is conveyed very well by this book's high-quality discussions.

Thus I recommend the book wholeheartedly not just to any philosopher of physics, but to any metaphysician and epistemologist who is minded to have their views moulded by the deliverances of empirical enquiry. All future work on the Everett interpretation begins here.

*Acknowledgements*: I am very grateful to Matthew Donald, Anthony O'Hear, Adrian Kent, Simon Saunders and David Wallace for comments on earlier drafts: if only I could have consistently followed all suggestions!

REFERENCES

Bell, J. (1986), 'Six possible worlds of quantum mechanics', *Proceedings of the Nobel Symposium 65* (Stockholm August1986); reprinted in Bell (1987), page references to reprint.

Bell, J. (1987), *Speakable and Unspeakable in Quantum Mechanics*, Cambridge University Press; second edition 2004, with an introduction by Alain Aspect.